\documentclass[twocolumn,prb,amsmath,floatfix]{revtex4}
\usepackage{graphicx}

\begin{document}

\title{Correlated and uncorrelated nanoscale heterogeneities in $L1_0$ solid solutions and their
signatures from local and extended probes}

\author{Rafael C. Howell}
\email{rhowell@lanl.gov}
\affiliation{Materials Science and Technology Division, Los Alamos National Laboratory,
Los Alamos, NM 87545, USA}

\author{Steven D. Conradson}
\email{conradson@lanl.gov}
\affiliation{Materials Science and Technology Division, Los Alamos National Laboratory,
Los Alamos, NM 87545, USA}

\author{Angel J. Garcia-Adeva}
\email{wuzgaada@lg.ehu.es}
\affiliation{Departamento de Fisica Aplicada I, E. T. S. de Ingenier\'{i}a de Bilbao, UPV/EHU,
Alda. Urquijo s/n, 48013, Bilbao, Spain}

\date{\today}

\begin{abstract}
The phase coexistence of chemically ordered $L1_0$ and chemically disordered structures within
binary alloys is investigated, using the NiMn system as an example.  Theoretical and numerical
predictions of the signatures one might expect in data from local and extended probes are
presented, in an attempt to explain the presence of antiferromagnetism in NiMn when no $L1_0$
signatures appear in diffraction data.  Two scenarios are considered, the first in which the
tetragonal $L1_0$ structure and fcc chemically disordered structure are distributed evenly into
uncorrelated domains of specified average diameter.  The diffraction limit, below which the two
structures can only be distinguished using a local probe, is quantified with respect to the domain
diameter by applying straightforward diffraction analysis.  In the second scenario, domains with
chemical ordering oriented in different directions are required to maintain their atomic coherence
with each other.  A numerical treatment is used to illustrate the long-range strain that results
from elastic energy considerations, and the effects on the structure factor (extended probe) and
pair distribution function (local probe) are investigated.

\centering LA-UR 06-2479
\end{abstract}

\maketitle

\section{Introduction}

The $M_xF_{1-x}$ intermetallics, where a metal, M, is combined with F, one of the three
ferromagnetic transition metals iron, nickel, or cobalt, for x close to 0.5 or 0.75, provide some
of the best known examples of the importance of chemical ordering as determinants of
structure:property relationships.  When crystallographically disordered by being prepared as, e.g.,
thin films deposited on low temperature substrates, they are paramagnetic and cubic. However, when
they are allowed to develop their most stable structures by preparing or annealing them at higher
temperatures, the different elements order as homogeneous layers occupying the [100] planes that
alternate with each other according to the overall stoichiometry of the material. They are then
commonly referred to as $L1_0$ alloys because the different sizes of their constituent elements as
well as the chemical ordering results in a tetragonal distortion perpendicular to the layers. The
proximity and periodicity of the atoms now cause the spins on the atoms of the FM metal to order
first within a layer and next between the neighboring layers so that the $L1_0$ alloys are ferro or
antiferromagnets, with the magnetization also along the tetragonal axis.  Of obvious interest are
the intermediate regime between the fully disordered and ordered materials, the path connecting
them, and the relationship between the partial structural spin ordering.

In the case of Ni$_x$Mn$_{1-x}$ alloy films near the equiatomic composition $x=0.5$, the magnetic
and structural properties have generated interest both from the basic and applied viewpoints.
Such films exhibit the aforementioned $L1_0$ tetragonal phase which comprises alternating atomic
planes of Ni and Mn along the $c$-axis with a contraction perpendicular to these planes.  In its
chemically ordered state, NiMn is a strong antiferromagnet, which makes it a good candidate for
use as the exchange coupling material to pin the reference ferromagnetic layer in spin valve
magnetic read sensors.\cite{nogues, devasahayan}  Chemically ordered NiMn is capable of providing
strong pinning fields, a high blocking temperature, and good corrosion
resistance.\cite{lin, mao, schonfeld, anderson, zhang}

It has been shown that the exchange bias in NiMn alloys decreases with decreasing antiferromagnet
layer thickness and that a critical thickness is needed to develop the exchange coupling.  It has
been suggested\cite{toney} that such thin films ($\leq5-10$ nm) are not chemically ordered and
hence are not antiferromagnetic, which explains the small exchange coupling in these films.  In
grazing incidence X-Ray diffraction measurements up to $Q\sim 5$ \AA$^{-1}$ on NiMn films annealed
at $280^o$ C over this range of thicknesses, no traces of sites were found.  In $15-25$ nm thick
films, the chemically ordered $L1_0$ phase grows at the expense of the fcc phase, increasing the
exchange bias to $300-400$ Oe.

An intriguing aspect of these $M_xF_{1-x}$ intermetallics is that the onset of magnetization
occurs prior to the observation of chemical ordering/tetragonal distortion. This behavior suggests
that the spins can align in magnetic domains that are at or below the size of structural domains
that would give signatures in diffraction, i.e., the diffraction limit. Local structure
measurements on PtCo, PtFe, and NiMn have in fact demonstrated a correlation between this initial
magnetization and ordering on local length-scales.\cite{tyson, cross}  It is notable that a very
high degree of local ordering was observed in these studies that suggests the materials are
actually composed of incoherent nanoscale domains that are close to the homogeneous $L1_0$
structure.

XAFS studies provide complementary information not only to diffraction data, which gives the
long-range average order, but also to that obtained from diffuse scattering and pair distribution
analysis, providing an alternative means for probing the development of the nucleation sites for
the $L1_0$ phase.\cite{warren1, lawson}  Indeed, XAFS has been employed at the Mn {\it K}-edge as
a direct and element specific probe of the average local atomic structure within
$\sim 5-6$ \AA\, of the Mn site, in which a short-range order parameter is introduced to quantify
and confirm the presence of $L1_0$ structures.\cite{espinosa}  At low annealing temperatures, the
short-range order parameter differs from an associated long-range order parameter obtained from
diffraction data, providing a sufficient condition for nanoscale phase separation and
heterogeneity by demonstrating the presence of a nondiffracting but locally ordered component
within the alloy.  In these films a nonzero coercivity is measured even while the associated
diffraction patterns exhibit no signs of an $L1_0$ structure.

When correlations between local distortions produce ordered distributions, they are generally 
observed in diffraction experiments as superlattice or satellite peaks (in addition to the Bragg 
peaks).\cite{withers}  The situation, however, is more complicated when domains below the limit of
long-range order occur in a random or aperiodic distribution.\cite{warren2, welberry1, welberry2}
In this case, the effects of heterogeneity, defect or other chemical organization, and competition
between different phases on the diffraction patterns are not well understood.  In
Garcia-Adeva {\em et al.},\cite{garcia-adeva1} for example, it has been shown that a second
coexisting structure affecting one-fifth or more of the atoms in various two dimensional lattices
can have a minimal to negligible effect on the calculated structure factors when compared with
their periodic counterparts.  The signatures of disorder that are present in the diffuse part of
the scattering are usually smooth and subtle, and it is often easy to construct several distinct
disordered lattices that produce essentially equivalent diffraction patterns.

The indications of chemical ordering within very small domains of sizes that will average
crystallographically to a random, cubic structure is not surprising.  Although even a random
distribution of atoms may produce a plethora of semi-ordered domains containing up to several
dozen atoms, given the stability of the layered structures in this system it is likely that when
new atoms are deposited they add to existing structures in an ordered way that is terminated when
the growing layered nanocrystallites collide. An alternative description might be to consider a
random solid solution where each atom is allowed to exchange its position with a neighbor to form
the more stable ordered structure only once, resulting in a large number of small domains that
individually exhibit a high degree of order internally. Because of the significance of this
phenomenon whereby various properties have size thresholds for their expression in this and other
systems, it is worth exploring their structural properties so as to understand the experimental
signatures of nanoscale ordering in crystalline materials and the ramifications of ordered domains
on stress and defect formation.\cite{garcia-adeva2}

Here, using the example of the NiMn system, we investigate the effects that domain size and
interface behavior have on conventional diffraction when nanoscale phase separation and
heterogeneity either occur below such a diffraction limit, or are faced with the restriction of
having coherent atoms across their respective interfaces in which long-range elastic strain
prohibits their preferred structures from forming within the domains.  We study the relevant
length-scales in which chemical order can exist in the form of small nanoscale $L1_0$ domains
immersed in a chemically disordered matrix with an average, crystallographic structure, and
quantify the domain sizes by which such distortions will not be apparent in the diffraction data
and yet still resolvable with local probes.  In addition, at this length-scale, we discuss the
role that elastic stresses at the domain interfaces play in determining the relaxation field of
the structure within the domains themselves, the energy considerations of such interfacial
interactions, and the associated signatures such a strain has on the long-range (e.g. diffraction)
and local (e.g. XAFS, pair distribution) probes.

The analytical and numerical analysis that follows will make use of the structure factor $S(Q)$
and pair distribution function (PDF) $G(r)$ of various lattices to illustrate the effects of
nanoscale heterogeneity on extended and local probe data, as the two functions are well-defined
and computationally tractable.  The formalism used is identical to Peterson
{\em et al.},\cite{peterson} with comparisons to other definitions and nomenclature found in
Keen \cite{keen}.

Nanoscale heterogeneity and phase separation are not exclusive to the magnetic alloy considered in
here.  This is also the case, for example, with substitutional impurities in solid
solutions.\cite{ice}  If the impurity atoms attract or repel each other, heterogeneous nanoscale
domains or texture will form in the solid.  Likewise, these effects are seen in charge separation
and charge ordering in colossal magnetoresistance compounds and high $T_c$ superconductors,
leading to the formation of stripes.\cite{xu, zaanen, moreo, dagotto, tranquada1, tranquada2}

\section{Intrinsic nanoscale ordering and the formation of correlated and uncorrelated domains}

A random arrangement of atoms will exhibit chemical ordering on all length-scales, with more
ordering appearing more often at smaller, nanoscale lengths.  In addition, a growth process might
favor the formation of such domains beyond their intrinsic random existence, if the energetics
permits.  The growth of layers during atom deposition, or neighbor exchange on a surface, are
examples mentioned before.  

\begin{figure}[ht]
\centering
\includegraphics[width=3in]{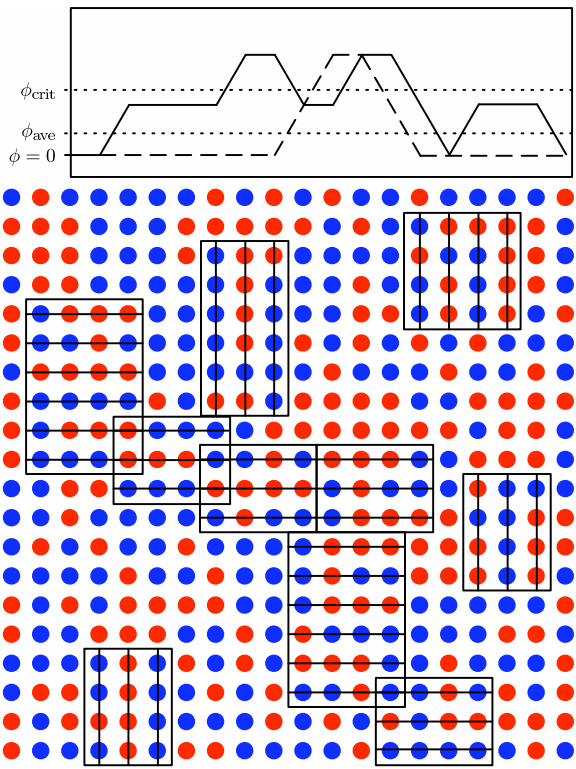}
\caption{A random arrangement of red and blue atoms will always display some ordering, depicted as
rectangles in this figure.  In certain cases, such ordering can trigger the onset of a function,
like antiferromagnetism.  This is illustrated with the use of an order parameter $\phi({\bf x})$
that attains some critical value within the cores of ordered domains.}
\label{random_ab}
\end{figure}

In the case of the $L1_0$ ordering just described, one might define a region, or domain, of such
ordering if more than 3/4 of the atoms satisfy the definition of $L1_0$ ordering, with alternating
rows of distinct atoms.  Figure \ref{random_ab} shows a two dimensional example of this, in which
equal numbers of red and blue atoms are distributed randomly.  Rectangular regions are demarcated
if their enclosed atoms show such ordering, and the parallel lines within the domains show the
$L1_0$ orientation.  Using these rules, it is seen that roughly $35-40\%$ of the atoms are found
in such layered-type domains.  With this, one can define an order parameter $\phi({\bf x})$ that
has a value of zero in locations where no $L1_0$ ordering is present, a value of one within
domains exhibiting $L1_0$ ordering, and a value of one-half at the interfaces.  The top of the
Fig. \ref{random_ab} shows a plot of $\phi$ across the middle horizontal row of atoms, in which
the ordering is high (solid line), and a horizontal row from a lower portion of atoms, in which
the ordering is low (dashed line), along with values of the average ordering $\phi_{\rm ave}$ and
a hypothetical critical value $\phi_{\rm crit}$.  When the ordering is above the critical value,
the structure takes on a function, in this case antiferromagnetism.  Recent calculations of local
electronic density of states in NiMn do in fact show that the perturbation on the spin ordering
from the inhomogeneity at the interface is spatially confined.\cite{garcia-adeva2}  The plot shows
that indeed certain regions attain this value, despite the random arrangement of atoms.  The
majority of ordering will exist at the nanoscale, and hereafter we set out to determine the
signatures that such ordering exhibit in extended and local probes.

Our treatment of the fcc-$L1_0$ phase coexistence in NiMn alloys will mostly involve cases in
which the two structures are present in equal amounts, and we consider the case of total $L1_0$
ordering for completeness.  We have found that the degree of atomic correlation at the interfaces
of domains plays a critical role in determining the actual crystallographic structure within the
domains.  In the case where there is no correlation whatsoever between the orientation and
placement of one domain with respect to another, then it is an excellent approximation to assume
that the internal structure of the domain is identical to that of an associated infinite crystal
(that is, it is ``ideal'').  Small distortions at the interfaces contribute insignificant
signatures to $S(Q)$ and $G(r)$.

For small domain sizes, the structure within the domains can be close to ideal if one allows for
dislocations at the interface between the domain and the matrix it is embedded within.
Dislocations serve to localize the large elastic energy present in any system containing two or
more phases with different lattice parameters between them.  Without dislocations, the internal
structures of the domains and the matrix would be far from ideal, exhibiting a continuous strain
throughout such that the structure remains coherent at the interfaces.

\begin{figure}[ht]
\centering
\includegraphics[width=3in]{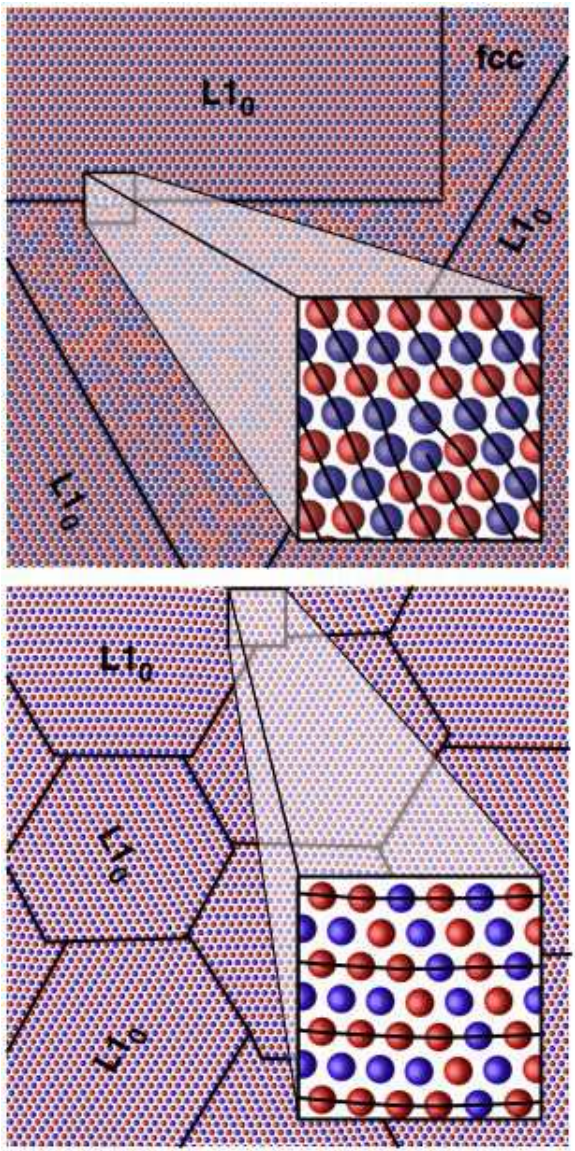}
\caption{Top: A [111] plane showing portions of three chemically ordered $L1_0$ domains separated
by a chemically disordered fcc phase.  The lattice mismatch is accounted for locally by the
presence of a dislocation (inset), which might allow the structure to attain the separate fcc and
tetragonal phases elsewhere.  Bottom: complete $L1_0$ ordering within hexagonal domains can force
the crystal into a highly strained state if no dislocations are allowed, thus diminishing the
presence of either fcc or tetragonal structures.}
\label{nimn_2d}
\end{figure}

The top image in Fig. \ref{nimn_2d} illustrates how a dislocation at the interface can localize
the strain such that the internal structures of $L1_0$ domains and the fcc matrix could take on
their ideal structures.  The [111] plane of a NiMn alloy with both $L1_0$ and fcc structures can
be described as a triangular arrangement of atoms, with the $L1_0$ (tetragonal) structure in this
plane having expanded atomic separations along rows of similar atoms and contracted atomic
separations along rows of different atoms.  The fcc structure has a higher symmetry chemically
disordered phase in which the average bond length between any two neighbor atoms is constant.  The
lattice mismatch between the structures is accounted for locally by the presence of a dislocation,
shown in the inset.  Although the dislocation introduces a large distortion locally in the
lattice, distant atoms are able to take on their preferred arrangement, even at the interface of
the $L1_0$ and fcc domains.

The bottom image in Fig. \ref{nimn_2d} depicts an arrangement of hexagonal domains with complete
$L1_0$ ordering, in which no dislocations are allowed.  The resulting stress on the atoms induces
a strain that propagates throughout the domains, as depicted in the inset.  The length-scale over
which the strain extends tends to alter the unique tetragonal structures that are otherwise
present in the case of the rectangular domains that allow for dislocations.  We will see that such
an arrangement of domains, and the accompanying strain, can actually produce a structure with an
increased tetragonal distortion, albeit highly disordered.

It is useful to consider whether the presence of dislocations, which localize the large elastic
energy around a core of atoms while allowing a minimum elastic energy configuration elsewhere, is
energetically favorable over the case of an identical distribution of domains without
dislocations.  The latter scenario tends to uniformly distribute the elastic energy throughout a
large region near the interfaces.  Consider a [111] plane of NiMn alloy in which an arrangement
of two dimensional $L1_0$ rectangular domains is embedded within a chemically disordered fcc
structure, as depicted in the upper half of Fig. \ref{nimn_2d}.  The total elastic energy is
minimized by varying the positions of the atoms (the exact interaction potentials between the
atoms are detailed in the fourth section).  In one case, dislocations are allowed at the
interfaces.  In the other case, the distribution of domains is identical, but no dislocations are
present and each row of atoms spans the entire lattice, thus maintaining coherence.

\begin{figure}[ht]
\centering
\includegraphics[width=3in]{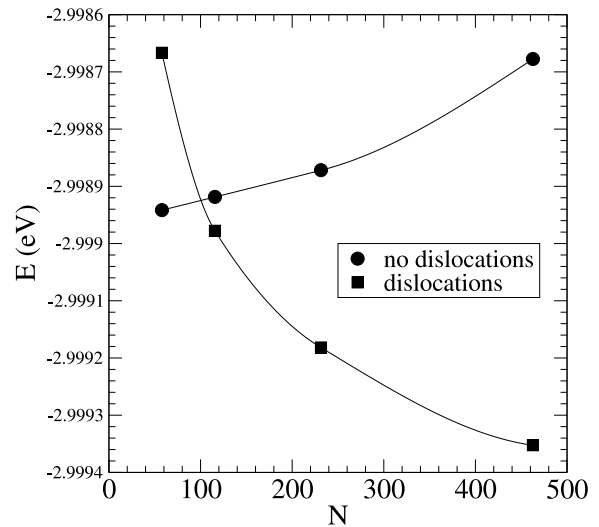}
\caption{The average energy per atom as a function of rectangular domain size, characterized by
the average number of atoms $N$ per side.  The lattice mismatch of the two structures determines
the domain size in which dislocations at the interfaces are favorable.}
\label{e_vs_d}
\end{figure}

Figure \ref{e_vs_d} shows the minimum elastic energy of uncorrelated (dislocations allowed,
incoherent) and correlated (no dislocations, coherent) domains for the two dimensional case.  The
rectangular domain size is characterized by the average number of atoms $N$ per side.  Clearly, it
becomes more favorable to have uncorrelated domains coexisting within the two dimensional
structure as the average domain size increases.  For $N\lesssim 100$, however, dislocations at the
interfaces introduce more elastic energy per atom than highly correlated structures in which the
strain is uniformly distributed throughout the interior of the domains.  In this case, the lattice
mismatch between the $L1_0$ and fcc structures is too small to be accounted for by the addition or
subtraction of one dislocation row.

This analysis of a two dimensional system motivates one to consider both scenarios.  Clearly,
correlated domain distributions are favorable when the domain size is small, and we will
investigate their effects on $S(Q)$ and $G(r)$ for various domain sizes.  In this case energetics
plays an important role, and the atom positions (and the ensuing long-range strain) must be
carefully accounted for with a robust numerical model.  In the next section, however, we consider
the case in which spherical domains of ideal $L1_0$ and fcc structures coexist in a three
dimensional structure.  By incorporating the reasonable approximation that the domains are
entirely uncorrelated with each other, through the presence of dislocations or otherwise, we
quantitatively obtain their diffraction and PDF signatures with little analytical effort, and
without any need for large-scale numerical treatments.

\section{The structure factor of three dimensional uncorrelated fcc and $L1_0$ spherical domains}

The radially averaged structure factor, $S(Q)$, of a NiMn alloy with uncorrelated fcc and $L1_0$
spherical domains can be derived analytically with very little approximation needed.  Consider a
NiMn alloy in which the fcc domains have an internal structure defined by a unit cell with lattice
parameter ${\bar a}=3.65$ \AA, and the $L1_0$ tetragonal domains have an internal structure
defined by a unit cell with $a=3.72$ \AA\, and $c=3.52$ \AA.\cite{hellwege}  The atomic densities
of the two internal structures are equal within $0.2\%$.

The structure factor can be obtained directly from the pair distribution function, $G(r)$, by a
Sine transform.  We have previously shown that uncorrelated spherical domains with a particular
crystal structure and average diameter $d$ generate a PDF that is simply that of the associated
infinite crystal multiplied by an envelope function that is characterized by $d$.\cite{howell}  As
a result, $S(Q)$ for the uncorrelated spherical domains is obtained from that of an infinite
crystal by a related convolution, that broadens the Bragg peaks, giving them a half-width of
approximately
\begin{equation}
\Delta Q=\frac{3.48}{d}.
\label{width}
\end{equation}

Consider the effect such a convolution has on the ideal Bragg peaks in the interval
$0\leq Q\leq5$ \AA$^{-1}$.  The conclusions made within this interval can easily be extended to
larger intervals.  The ideal fcc structure in the NiMn alloy has three Bragg peak positions at
$Q_1=2\sqrt{3}\pi/{\bar a}$, $Q_2=4\pi/{\bar a}$, and $Q_3=4\pi\sqrt{2}/{\bar a}$.  The ideal
$L1_0$ structure has five Bragg peaks at $Q_1$, $Q_2-\Delta Q_2^-$, $Q_2+\Delta Q_2^+$,
$Q_3-\Delta Q_3^-$, and $Q_3+\Delta Q_3^+$, where the splitting widths from the second and third
fcc peaks are given by
\begin{eqnarray}
\Delta Q_2^-&=&4\pi\left(\frac{1}{{\bar a}}-\frac{1}{a}\right)=0.0652 \text{ \AA}^{-1} \nonumber \\
\Delta Q_2^+&=&4\pi\left(\frac{1}{c}-\frac{1}{{\bar a}}\right)=0.127 \text{ \AA}^{-1} \nonumber \\
\Delta Q_3^-&=&4\sqrt{2}\pi\left(\frac{1}{{\bar a}}-\frac{1}{a}\right)=0.0922 \text{ \AA}^{-1} \nonumber \\
\Delta Q_3^+&=&4\pi\left(\sqrt{\frac{1}{a^2}+\frac{1}{c^2}}-\frac{\sqrt{2}}{{\bar a}}\right)=0.0454 \text{ \AA}^{-1}.
\label{3d_splitting}
\end{eqnarray}

In the limit of very large average domain diameters, the Bragg peaks from both structures are
easily resolved in $S(Q)$.  As the average domain size decreases the broadening of the peaks
introduces an overlap between neighboring peaks that eventually removes each Bragg peak maxima
from the $L1_0$ structure (as its peaks are smaller in magnitude than those of the fcc structure).
The peaks then become ``shoulders'' of their neighboring fcc peaks, and are still distinguishable
because of changes in the local curvature of the structure factor between the two peak locations
(i.e., an inflection point still resides between them).  Eventually, however, the average domain
diameter $d$ becomes small enough that such curvature disappears altogether, and the $L1_0$ peak
is no longer resolvable.

Such a ``diffraction limit'' for the domain diameter can be easily determined by equating the
convolution half-width given by Eq. \ref{width} with each of the splitting widths in
Eq. \ref{3d_splitting}.  The following domain diameters will achieve such a limit for the
associated $L1_0$ peak locations in the total structure factor (arranged by increasing domain
diameter):
\begin{eqnarray}
&&Q_2+\Delta Q_2^+: d=27.5 \text{ \AA} \nonumber \\
&&Q_3-\Delta Q_3^-: d=37.8 \text{ \AA} \nonumber \\
&&Q_2-\Delta Q_2^-: d=53.4 \text{ \AA} \nonumber \\
&&Q_3+\Delta Q_3^+: d=76.6 \text{ \AA}.
\label{diameters}
\end{eqnarray}

To illustrate the accuracy of such a simple analysis, consider the example of a NiMn crystal with
a normalized distribution of spherical particle diameters $d'$ (Howell {\em et al.}\cite{howell}
consider an entire family of distributions, but the characteristic peak broadening given by
Eq. \ref{width} is independent of the actual distribution),
\begin{equation}
P(d',d)=\frac{2}{3d}\left(4\frac{d'}{d}\right)^3 e^{-4\frac{d'}{d}},
\label{prob}
\end{equation}
with half the domains comprising an fcc structure and half an $L1_0$ structure.  The average
diameter of the domains is $d$ and the width of the distribution is $\sigma=d/2$.  In this
approximation, the regions between the spherical domains is a highly strained structure with no
long-range order, and the domains are entirely uncorrelated with each other.  The structure has a
constant local atomic density, approximately, within all the domains and the regions in between.

It can be shown that the PDF of such a structure is the average of the PDF of the two uncorrelated
domains, $G(r)=1/2[G_{fcc}(r)+G_{L1_0}(r)]f_e(r,d)$, where the envelope function for the
distribution of particle sizes given by Eq. \ref{prob} is $f_e(r,d)=(1+2r/d)\exp(-4r/d)$.  Note
that an ideal PDF constructed in this way still exhibits perfectly resolved pair distances for
$0\leq r\lesssim d$, so that a local probe such as XAFS should readily be able to determine the
presence of the two structures, and their relative amounts.

The structure factor is obtained from the PDF by a Sine transform, and is thus the average of the
structure factors of the two uncorrelated sets of domains.  The latter can be obtained by
convoluting the Cosine transform of the PDF envelope function with the structure factors of the
associated infinite crystals.  The expression for the total structure factor is then
\begin{widetext}
\begin{equation}
S(Q,d)=\frac{1}{2\pi Q}\int_0^{\infty}\left[{\bar f}_e(Q-Q',d)-
{\bar f}_e(Q+Q',d)\right]\left[S_{fcc}(Q')+S_{L1_0}(Q')\right]Q'\, dQ',
\end{equation}
\end{widetext}
where
\begin{equation}
{\bar f}_e(Q,d)=2d\frac{48+(Qd)^2}{\left[ 16+(Qd)^2\right]^2}
\label{conv}
\end{equation}
is the convolution function for the distribution given by Eq. \ref{prob}.  It has a Lorentzian-like
shape that serves to broaden the Bragg peaks in the structure factors of the ideal structures.

\begin{figure}[ht]
\centering
\includegraphics[width=3in]{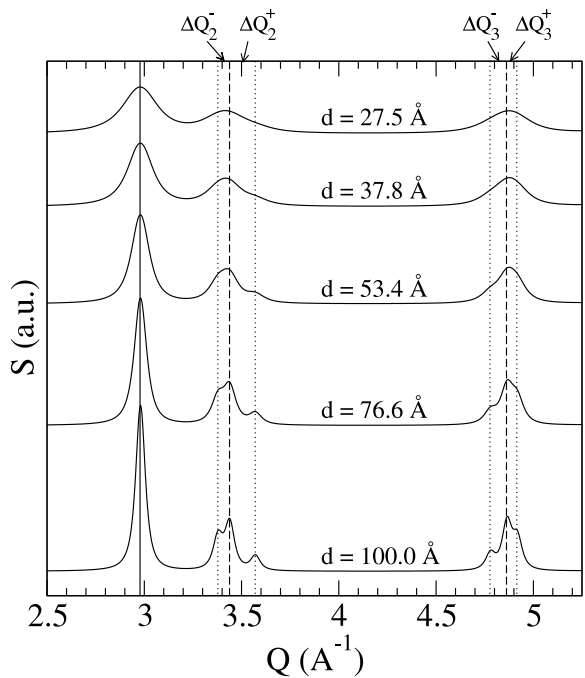}
\caption{The structure factor $S(Q)$ for a NiMn alloy with equal amounts of $L1_0$ and fcc 
structures within embedded domains with the average diameters shown.  The vertical
solid line is the first ideal Bragg peak position shared by the two coexisting structures.  The
two vertical dashed lines are the additional ideal peak positions for the fcc structure, and the
four vertical dotted lines are the additional ideal peak positions for the $L1_0$ structure. The
splitting widths, given by Eq. \ref{3d_splitting}, are labeled at the top.}
\label{sq_uc}
\end{figure}
   
Figure \ref{sq_uc} shows the structure factor of the NiMn alloy for each of the average diameters
given in Eq. \ref{diameters}, obtained by convoluting the ideal structure factor (seven delta
functions representing the ideal Bragg peaks of the fcc and $L1_0$ phases) with the function given
by Eq. \ref{conv}.  Also shown is the case when $d=100$ \AA, where all peaks are easily resolved.
The vertical lines are the locations of the ideal Bragg peak positions for the $L1_0$ and fcc
structures, and serve as guides to determine when individual peaks are resolvable.  Indeed, for
$d=76.6$ \AA, $53.4$ \AA, $37.8$ \AA, and $27.5$ \AA, the seventh, second, fifth, and fourth peaks
become entirely unresolved, respectively.  For all average domain sizes $d\lesssim 30.0$ \AA, the
total structure would be characterized as entirely fcc if one considered only the positions of
peak maxima and shoulders in $S(Q)$, while a local probe would readily resolve the two structures.

The diffraction limit for uncorrelated domains can be determined with little analytical effort,
and the domain size by which this limit is attained is quite small.  When domains coexist but
remain correlated, such that collections of atoms that span across domain interfaces maintain the
coherence of the crystal, then the stresses introduced by differing orientations of chemical
ordering can introduce significant strain throughout the domains, and such analysis is no longer
feasible.  For this scenario, we consider a numerical treatment to investigate the relationship
between domain size and diffraction limit.

\section{The structure factor and pair distribution function of two dimensional correlated fcc and
$L1_0$ rectangular domains}

The two dimensional lattices represent [111] planes of the NiMn alloy films discussed, within
which the $L1_0$ structure is comprised of alternating rows of the two elements.  In one class of
simulations, rectangular domains of various sizes of the $L1_0$ structure are embedded in a
chemically disordered triangular (fcc [111]) matrix, with each structure present in equal amounts.
Rectangles are constructed with N rows and N columns of atoms, giving an average side of length
${\bar d}=N(2.63$ \AA $+\sqrt{3}/2\cdot 2.56$ \AA$)/2$.  Another class of simulations considers
hexagonally arranged domains of $L1_0$ chemical ordering that populate the entire crystal (no
disordered fcc component), with a diameter $d$ that measures the span between opposing hexagonal
vertices.

Atomic interactions described with a Rose potential \cite{rose} determine the precise atom
positions required to minimize the total elastic energy at zero pressure.  The equilibrium bond
lengths are inferred from the observed three dimensional fcc and tetragonal lattice parameters
detailed in the last section: $a=$ Ni-Ni = Mn-Mn = 2.63 \AA\, and $b=$ Ni-Mn = 2.56 \AA, and a
cutoff distance of 3.60 \AA\, limits the interactions to nearest neighbors.  The chemically
disordered phase averages to an fcc [111] structure with average bond length of
${\bar a}=2.594$ \AA.  The bond energies and moduli are equal for all three interactions, thus
simplifying the model significantly while still capturing the relevant physics.  The lattices are
constructed with approximately one million atoms, and periodic boundary conditions are used in all
calculations.  Large lattice sizes are required to obtain the desired resolution in the
diffraction data ($\delta q=0.002$ \AA$^{-1}$), and as such three-dimensional simulations have
exceeded our current computational limits (although preliminary results indicate that our
conclusions will remain qualitatively the same).  Finally, our results given here are averages
obtained from ten lattices, each differing in their random arrangement of domains, but statistical
differences in their structure factors are small.  The ability to obtain a minimum elastic energy
within such large lattices allows us to demonstrate the importance of the elastic response of the
lattice.

The lattices are distinguished by the domain sizes of the $L1_0$ structure.  Regions of typical
lattices are depicted in Fig. \ref{nimn_2d}.  For the rectangular domains, portions of three
domains show the possible orientations of the $L1_0$ [111] chemical ordering as well as the
chemically disordered fcc structure that separates them.  We compare the cases in which no
dislocations can form at the interfaces between these phases, by maintaining the interactions of
each pair of atoms that is determined by the perfect fcc arrangement of the atoms prior to the
energy minimization routine, to those in which dislocations are allowed by continually allowing
for changes to the pairing of atomic interactions based on new nearest neighbor arrangements that
occur during the crystal relaxations. The hexagonal domains represent the extreme case in which
the crystal is entirely chemically ordered, and no dislocations can occur at any interfaces, such
that the preferred tetragonal configurations within each domain present significant stresses at
these interfaces.  One therefore expects this case to represent the limit to which a large degree
of related strain can occur throughout the crystal.  In all cases, because the different domain
orientations are present in equal amounts, the material does not exhibit any gross anisotropic
distortion, and thus any changes in the average crystal length along both directions are equal.

In the interval $0\leq Q\leq 5$ \AA$^{-1}$, the ideal triangular structure in the [111] plane of
the NiMn alloy has three Bragg peak positions at
$Q_1=4\pi/\sqrt{3}{\bar a}$, $Q_2=4\pi/{\bar a}$, and $Q_3=8\pi/\sqrt{3}{\bar a}$.  The ideal
$L1_0$ structure has seven Bragg peaks at $Q_1$, $Q_2-\Delta Q_2^-$, $Q_2$, $Q_2+\Delta Q_2^+$,
$Q_3-\Delta Q_3^-$, $Q_3$, and $Q_3+\Delta Q_3^+$, where the splitting widths from the second and
third peaks are given by
\begin{eqnarray}
\Delta Q_2^-&=&\frac{4\pi}{\sqrt{3}}\left(\frac{\sqrt{3}}{{\bar a}}-\sqrt{\frac{4}{a^2}-\frac{2}{ab}+\frac{1}{b^2}}\right)=0.0661 \text{ \AA}^{-1} \nonumber \\
\Delta Q_2^+&=&\frac{4\pi}{\sqrt{3}}\left(\sqrt{\frac{1}{a^2}-\frac{2}{ab}+\frac{4}{b^2}}-\frac{\sqrt{3}}{{\bar a}}\right)=0.0646 \text{ \AA}^{-1} \nonumber \\
\Delta Q_3^-&=&\frac{8\pi}{\sqrt{3}}\left(\frac{1}{{\bar a}}-\frac{1}{a}\right)=0.0770 \text{ \AA}^{-1} \nonumber \\
\Delta Q_3^+&=&\frac{8\pi}{\sqrt{3}}\left(\frac{1}{b}-\frac{1}{{\bar a}}\right)=0.0739 \text{ \AA}^{-1}.
\label{2d_splitting}
\end{eqnarray}

Assuming that Eq. \ref{width} might apply in two dimensions as well, we can at once conclude that
if the two coexisting structures were entirely uncorrelated with each other, then when the average
domain side lengths are larger than ${\bar d}=3.48/\Delta Q_2^+=53.9$ \AA, all of the structure
factor peaks would be easily resolved. When they become smaller than
${\bar d}=3.48/\Delta Q_3^-=45.2$ \AA, all peaks from the $L1_0$ [111] structure would be
unresolved and the structure factor would indicate that only an fcc [111] was present.  Note the
exceedingly small domain size interval between these two limits, which is only a consequence of
considering two dimensional structures.

Correlated structures preclude the use of such simple analysis, however, as the crystal structure
within the $L1_0$ chemically ordered domains do not exhibit the tetragonal distortions seen in
their uncorrelated counterparts.  Requiring that every row in the two dimensional structure be
coherent across the entire lattice introduces a significant energy constraint into the system.  We
shall see that the behavior of the resulting strain is somewhat nonintuitive, by consistently
changing the ordered $L1_0$ structure into something very different than that suggested by the
pair interaction potentials alone.  Indeed, considering the two cases (50\% and 100\% $L1_0$
ordering) offers two extreme consequences of such strain.

\begin{figure}[ht]
\centering
\includegraphics[width=3in]{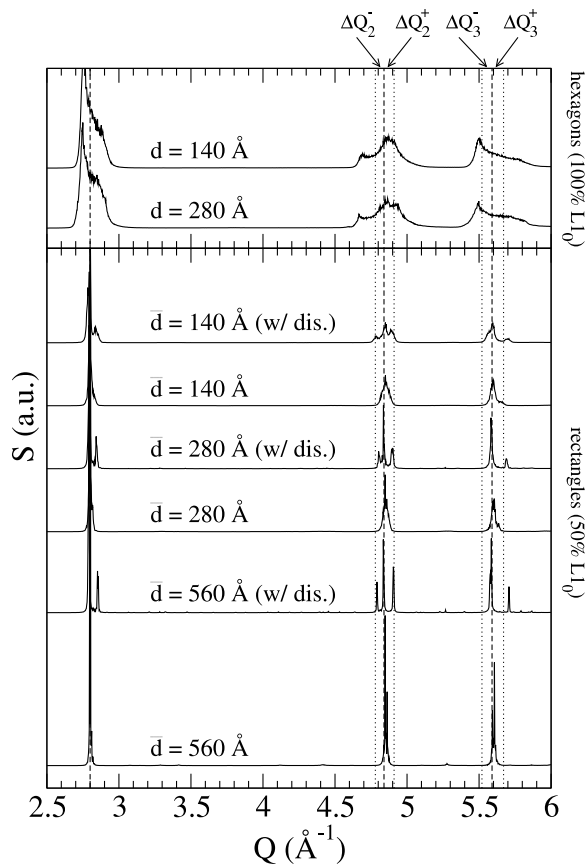}
\caption{The structure factor $S(Q)$, in arbitrary units and offset for comparison, for two
dimensional [111] planes of NiMn alloy.  Bottom: the structure factor from lattices with equal
amounts of fcc and $L1_0$ structures, the latter in embedded rectangular domains with the average
side lengths shown.  Top:  the structure factor from lattices entirely populated with hexagonal
$L1_0$ ordered domains, scaled by a factor of seven for comparison.  The three vertical dashed
lines are the ideal Bragg peak positions shared by fcc and $L1_0$ structures.  The four vertical
dotted lines are the additional ideal peak positions for the $L1_0$ tetragonal structure.  The
splitting widths are labeled at the top.}
\label{sq_c}
\end{figure}

Figure \ref{sq_c} shows the structure factors of the NiMn alloys for the correlated rectangular
domains (bottom) with sides of length ${\bar d}=560$ \AA, $280$ \AA, and $140$ \AA, including the
cases in which dislocations are allowed.  In all correlated cases (no dislocations), the second,
fourth, and fifth Bragg peaks for the tetragonal structure are never present (the vertical lines
in the figure represent the ideal Bragg peak positions of both structures).  Considering the
analysis of the three dimensional structures in the previous section, it is clear that their
absence is not strictly a matter of the structure having attained the diffraction limit.  Instead,
the tetragonal structure is not able to form entirely within the domains due to the coherence
condition at their interfaces, and thus the structure factor is revealing an average structure of
the fcc and $L1_0$ structures.  When dislocations are present, the tetragonal structure is easily
resolved at these domain sizes, and the peaks show a trend consistent with their three dimensional
counterparts, in which decreasing domain sizes diminishes the peak heights at the expense of
increased peak width.  One cannot conclude, however, if any fcc structure is present in such
systems, as the fcc Bragg peaks coincide with those of the tetragonal structure ($Q_1$, $Q_2$, and
$Q_3$) by construction.  A real-space treatment, utilizing the pair distribution function $G(r)$
will address this issue.

The upper portion of Fig. \ref{sq_c} shows $S(Q)$ for the complete chemically ordered crystals
with correlated hexagonal domains.  The peak structure is consistent with that of a single
mesoscale glassy phase, with extreme peak broadening due to short length-scale structural ordering.
In addition, significant shifts of the peak positions from the tetragonal phase, compared to
structures with rectangular domains, indicate the presence of a large tetragonal distortion.
Indeed, the hexagons in Fig. \ref{nimn_2d} are oriented in a consistent manner such that no two
neighboring domains share the same orientation, which promotes tetragonal distortions far larger
than those dictated by the interaction potentials.  This is particularly noticeable from the
shifts in the peak positions about $Q_2$ and $Q_3$.

\begin{figure}[ht]
\centering
\includegraphics[width=3in]{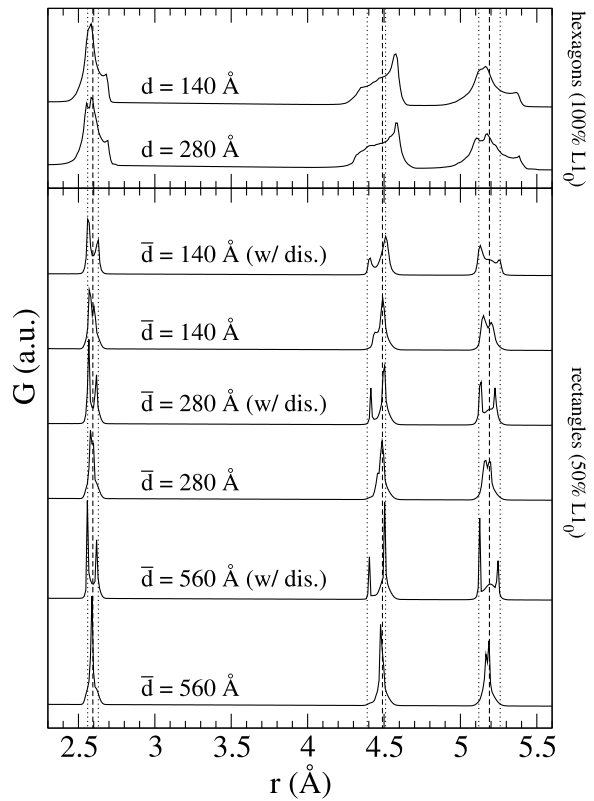}
\caption{The pair distribution function $G(r)$, in arbitrary units and offset for comparison, for
the same lattices represented in Fig. \ref{sq_c}.  The three vertical dashed lines denote are the
ideal pair distances for the fcc structure, while the six vertical dotted lines are for pair
distances of the tetragonal structure.  $G(r)$ from lattices entirely populated with hexagonal
$L1_0$ ordered domains are scaled by a factor of two for comparison.}
\label{gr_c}
\end{figure}

To better understand the structures, we show in Fig. \ref{gr_c} the pair distribution function
$G(r)$ for the same relaxed lattices, with the dashed vertical lines denoting the ideal pair
distances for the fcc structure and the vertical dotted lines for those of the tetragonal
structure.  Interestingly, when dislocations are allowed at the rectangular interfaces (bottom),
the entire lattice seems to relax to the tetragonal structure, even though the $L1_0$ chemical
ordering exists in only half the crystal.  Only a very low amplitude peak is seen at the third fcc
pair distance in these cases.  This may be a result of collective behavior among the atoms within
the domains, in which the ordering allows for a common, global interaction-like scenario that
easily influences the random interactions outside the domains.  When coherence across the
interfaces is enforced, however, the distribution of lengths from the first through third nearest
neighbors generally indicates the presence of only one average fcc structure, with tetragonal peak
splittings being revealed only in the smallest domain size considered.

The lattices with entire hexagonal domains exhibit extremely broad and shifted peaks in the pair
distribution function, a result of significant and continuous tetragonal distortions.  The peak
structure present in $G(r)$ decays away to a uniform atomic density at approximately
$r=25$ \AA\, (not shown in Fig. \ref{gr_c}), which suggests a description of the structure as a
mesoscale glass with medium-range order.  Additionally, the peak positions, when resolvable, are
shifted from the $L1_0$ tetragonal structure so much that one can only conclude that some
collective behavior among the domains is present, which we alluded to earlier with respect to
their structure factors.

When there is only one average structure present throughout the crystal, such as the case with
correlated rectangular domains of $L1_0$ chemical ordering comprising half the structure, then
local and extended probes reveal the same structure.  Then, in the context of the NiMn alloy
considered here, magnetic order is probably more correlated with the $L1_0$ chemical ordering than
the tetragonal distortions.

\section{Conclusion}

We have provided conclusive evidence, both analytically and numerically, that it is indeed
possible to have well-defined tetragonal distortions within localized domains that fall below
the diffraction limit.  The structure is therefore readily attainable via local probes such as XAFS.
An analytical treatment considered the phase coexistence of such structures with a chemically
disordered fcc structure, where the two structures are entirely uncorrelated.  Basic diffraction
analysis predicts with great accuracy the domain size below which the tetragonal signatures are no
longer present in unique Bragg peaks, but instead become part of the diffuse scattering of the
average fcc structure.

In the case in which atomic coherence is maintained across the interfaces of domains, numerical
treatments incorporating elastic energy considerations indicate that the tetragonal structure
may only manifest itself far away from the interfaces, but the domain sizes considered in this
treatment never allowed for true fcc and tetragonal phase separation.  Instead, the strain incurred
by the competing domains is of such great extent as to produce even larger tetragonal distortions
than the pair interaction potentials dictate, as well as significant disorder indicative of
mesoscale glassy materials (seen in the 100\% $L1_0$ ordered hexagonal domains).  Alternatively,
it has a tendency to average the structure to an fcc phase (as in the case of 50\% $L1_0$ ordered
rectangular domains).

If the latter case is indeed the correct description of the nanoscale ordering of NiMn alloys,
then one must conclude that the presence of antiferromagnetism is a direct result of the chemical
ordering present in the $L1_0$ structure, and not the actual tetragonal distortion since this is
suppressed.  This type of smooth transition in the bond lengths is however what was observed
experimentally in the NiMn system.\cite{espinosa}

RCH and SDC acknowledge supporte from the Heavy Element Chemistry Program, Chemical Sciences,
Biosciences, and Geosciences Division, Office of Basic Energy Sciences, and Defense Programs,
NNSA, U.S. Department of Energy under Contract No. W-7405.  AJG-A acknowledges financial support
from the Spanish MEC under a Ramon y Cajal contract.

\end{document}